\documentclass[pra,showpacs,preprintnumbers,floatfix,superscriptaddress,10pt]{revtex4}
\usepackage[mathscr]{eucal}
\usepackage{color}
\usepackage[dvips]{graphicx}
\usepackage{epsf}
\usepackage{bm}

\newcommand{\be}{\begin{equation}}
\newcommand{\ee}{\end{equation}}
\newcommand{\ba}{\begin{eqnarray}}
\newcommand{\ea}{\end{eqnarray}}

\begin{document}

\title{Bulk viscosities for cold Fermi superfluids  close to the unitary limit}

\author{Miguel Angel Escobedo}
\email{mesco@ecm.ub.es}
\affiliation{
Departament d'Estructura i Constituents de la Mat\`eria and \\
Institut de Ci\`encies del Cosmos, Universitat de Barcelona \\
Diagonal 647, E-08028 Barcelona, Catalonia, Spain }

\author{Massimo Mannarelli}
\email{massimo@ieec.uab.es}
\affiliation{Instituto de Ciencias del Espacio (IEEC/CSIC) Campus Universitat Aut\`onoma de Barcelona, Facultat de Ci\`encies, Torre C5, E-08193 Bellaterra (Barcelona), Catalonia, Spain}

\author{Cristina Manuel}
\email{cmanuel@ieec.uab.es}
\affiliation{Instituto de Ciencias del Espacio (IEEC/CSIC) Campus Universitat Aut\`onoma de Barcelona, Facultat de Ci\`encies, Torre C5, E-08193 Bellaterra (Barcelona), Catalonia, Spain}
\preprint{UB-ECM-PF 09/12}

\begin{abstract}
We compute the coefficients of  bulk viscosity for a non-relativistic superfluid corresponding to a fermionic system close  to the unitarity limit. We  consider the  low temperature regime assuming that the transport  properties of the system are dominated by phonons. To compute the coefficients of bulk viscosity we use kinetic theory in the relaxation time approximation and the low energy effective field theory
of the corresponding system.  We show that the three independent  bulk viscosity coefficients, $\zeta_1, \zeta_2, \zeta_3$,  associated with irreversible flows vanish for  phonons with a  linear dispersion law. Considering  a phonon  dispersion law with a cubic term in momentum we find that in the conformal limit $\zeta_1 = \zeta_2=0$,  while $\zeta_3$ is non-zero.  
Including a conformal breaking term which arises for a large but finite $s$-wave scattering length, $a$, at the leading order in  $1/a$ we obtain that   $\zeta_1 \propto 1/a$ and  $\zeta_2 \propto 1/a^2$.  
 \end{abstract}

\date{\today}
\maketitle
\section{Introduction}

The properties of quantum degenerate fermionic systems with an attractive two-body scattering interaction  have been the subject of extensive investigation in the last years~\cite{Giorgini:2008zz}. Of particular interest are systems with an infinite two-body scattering length that are believed to have universal properties~\cite{Ho:2004zza}, meaning that the features of the system are independent of the detailed form of the inter-particle potential.   

Experiments with trapped cold atomic gases are able to reach the region of infinite scattering length (the so-called unitarity limit) tuning the interaction between the fermionic atoms by means of  a magnetic-field Feshbach resonance \cite{Ohara:2002,Bourdel:2003zz,Gupta:2003,Regal:2003}.  In general, in these experiments  the two different  populations of fermions consist of  atoms, like $^6{\rm Li}$ or $^{40} {\rm K}$, in two hyperfine states. 
The strength of the interaction between atoms  depends on  the applied  magnetic field and can be  measured  in terms of the $s$-wave scattering length. By varying the magnetic-field controlled interaction,  fermionic pairing is observed to undergo the Bose-Einstein condensate (BEC) to Bardeen-Cooper-Schrieffer (BCS) crossover. The unitary limit is reached when the magnetic field is tuned at the Feshbach resonance~\cite{Feshbach}, where the  two-body scattering length  diverges.

Far from the unitarity limit the properties of the   system  are qualitatively well understood
using mean field theory~\cite{BEC}.  In the weak coupling BCS region  the system is characterized by the formation of Cooper pairs. In the  strong coupling limit the system can be described as a  BEC dilute gas. 
The extreme BCS and BEC regimes are also in good quantitative control in mean field theory. 
However, the mean field expansion is not reliable close to  unitarity because the scattering length is much larger than the inter-particle distance and  there is no small parameter in the Lagrangian to expand in. Therefore  fluctuations may change the mean field results substantially. 

Close to the unitarity region a quantitative understanding of the phases comes mainly from
Monte-Carlo simulations~\cite{Carlson:2005kg}. Other approaches consist in considering the expansion in a  small parameter that comes from the generalization to an arbitrary number $N$ of spins~\cite{Radzihovsky:2007}, or to $d$ space dimensions~\cite{Nishida:2006br}. In the former case,   for $N \to \infty$ the problem is exactly solvable by mean-field theory and one can consider $1/N$ corrections   and then extrapolates to $N=2$.   In the second case one considers  an $\epsilon= 4-d$ expansion and then extrapolates the result to $d=3$. Both approaches are  in  quantitative agreement with Monte-Carlo simulations for
the equation of state of the system.
 A different method consists in adding the quantum fluctuations on the top of  the mean-field theory~\cite{Hu:2006,Randeria:2007}. In this way one improves the agreement with Monte-Carlo simulations, with respect to the bare mean-field calculation.
 
 It is a remarkable aspect  of these fermionic  systems that  for any value of the  attractive interaction  they are  superfluid, provided the temperature is sufficiently low.
Superfluidity is a phenomenon that occurs after the appearance of a quantum condensate that breaks a global $U(1)$ symmetry of the system~\cite{landaufluids,IntroSupe,Griffin}, regardless of whether the system is  fermionic or bosonic. In both cases   
Goldstone's theorem predicts the existence of low energy modes with a linear dispersion law, which are essential to explain the
property of superfluidity. We will refer generically to these modes as superfluid phonons, or phonons for simplicity, which
dominate the transport effects of the system at very low temperature.  
 
 The hydrodynamic equations governing the bulk fluctuations of  a superfluid are essentially different from standard fluid equations. At non-vanishing temperature one has to employ the two-fluid description of Landau
~\cite{landaufluids}, which takes into account  the motion of both the superfluid and of the normal component of the system.   
In order to describe the different dissipative processes one has to introduce more transport coefficients
than in a normal fluid. In particular, one has three independent bulk viscosities~\cite{footnotezetas}, $\zeta_1, \zeta_2, \zeta_3$, as well as the  shear viscosity and the thermal conductivity. 
The shear viscosity  of a unitary superfluid at  low temperature  has been computed in Ref.~\cite {Rupak:2007vp}. In Ref.~\cite{Son:2005tj}  it has been shown by a general argument that  $\zeta_1$ and  $\zeta_2$ vanish in the conformal  limit~\cite{Son:2005tj}. However, $\zeta_3$  and the thermal conductivity associated with irreversible heat flow cannot be determined by the same symmetry reasoning.  

In the present paper we evaluate the three bulk viscosity coefficients in the low temperature regime, $T \ll T_c$, where $T_c$ is the critical temperature for superfluidity, as we   consider the contribution of phonons only, assuming that the contribution of other degrees of freedom is thermally suppressed.   We show that all the bulk viscosities  vanish for phonons with a linear dispersion law. This is a result that was obtained several years ago  by Khalatnikov and Chernikova~\cite{Khalat-Cherni}, although not widely known. These authors also found that for superfluid $^4$He  all the transport coefficients, with the exception of the shear viscosity, vanish if one considers only phonons  with a linear dispersion law. 
Then, we evaluate the transport coefficients considering  a phonon  dispersion law that includes cubic 
corrections. We report the general expressions for the three dissipative coefficients,
given in terms of the parameters that appear in the phonon dispersion law and in terms of the pertinent decay rate.  Our results strongly depend on the coefficient of the cubic term in  the phonon dispersion law, which   is poorly known.   If  experiments and/or Monte-Carlo simulations could measure more precisely the value of this coefficient then we would know with  more accuracy the  numerical values of the bulk viscosities.

For the specific system we are interested in, the properties of the phonons needed in our computation
 can be extracted from the effective field theory constructed in Ref.~\cite{Son:2005rv}.
The Lagrangian of this theory   is  determined  by demanding  non-relativistic general coordinate invariance and conformal invariance and assuming  that phonons are the only relevant degrees of freedom.
From the effective Lagrangian  one  obtains  that the phonon dispersion law and self-couplings depend on some universal and dimensionless constants. Employing the expressions for these quantities we show that  in the conformal limit $\zeta_1 = \zeta_2=0$,  while $\zeta_3$  is  non-zero.

As a final step we study a system that is close to unitarity.  In this case scale invariance is broken  and additional terms in the effective Lagrangian must be taken into account~\cite{Son:2005rv}. 
In the presence of a large but finite  $s$-wave scattering length $a$,  we study  how the phonon dispersion law and the three-body self-couplings are modified.  Then, we evaluate  the first non-vanishing corrections to the bulk viscosity coefficients and  find that   $\zeta_1 \propto 1/a$ and  $\zeta_2 \propto 1/a^2$.  

It is also interesting to compare our results with those corresponding to a  Bose superfluid. A computation of all the transport coefficients
for a dilute but condensed Bose gas due to phonons was presented in Ref.~\cite{Kirkpatrick}, see also Ref.~\cite{Griffin}
for a more extended discussion. Remarkably, the temperature dependence for the bulk viscosities in these two different superfluids are the same.

Let us finally point out that the techniques we employ, and even the explicit computations, are very similar to those used in the evaluation of the   transport coefficients for  relativistic superfluids, in particular for the   color flavor locked  phase of dense quark matter~\cite{Manuel:2004iv,Manuel:2007pz}.
 
This paper is organized as follows. 
 In Section~\ref{Sec-hydro} we briefly review the superfluid hydrodynamics for a non-relativistic fluid including effects due to dissipation. The dissipative coefficients can be determined using kinetic theory and in Section~\ref{Sec-bulkviscosity} we give general expressions for the bulk viscosity coefficients  employing the two different methods described  in Ref.~\cite{IntroSupe}. In Section~\ref{Sec-lowenergy} we present the low energy effective theory valid for a cold Fermi gas,  in the exact unitarity limit, Sec.~\ref{Ex-EffeThe}, and close to it,  Sec.~\ref{Close-EffeThe}. The explicit evaluation of  the bulk viscosity coefficients is reported in Sec.~\ref{Sec-bulkviscosity-CA}. We draw our conclusions in Sec.~\ref{Sec-conclusion}.

Throughout, we use natural units, so that we take  the Boltzmann and Planck constants as $k_B = \hbar = 1$
in all our computations, but we restore them in the final section.

\section{Superfluid hydrodynamics and kinetic theory for the phonons}\label{Sec-hydro}

In this  Section we briefly review  the two-fluid theory of  superfluidity  developed by Landau.
We also provide the main ingredients of the kinetic theory for superfluid systems described by
 Khalatnikov~\cite{IntroSupe}.  This Section might be skipped
by the reader  who is familiar with superfluid hydrodynamics.

\subsection{Hydrodynamics of a non-relativistic superfluid }

In a superfluid there  are two  independent motions,
one normal and the other superfluid, with velocities ${\bf v}_n$ and ${\bf v}_s$, respectively. These motions  are associated with two different matter  and  current densities,  such that   the total density and the total current density  of the system are given by the sum of the superfluid and normal components as
\begin{equation}
\rho = \rho_n + \rho_s \ , \qquad {\bf j} =\rho_n {\bf v}_n + \rho_s {\bf v}_s \,.
\end{equation}

The superfluid motion is irrotational, ${\rm curl} \, {\bf v}_s = 0$, and thus its velocity
can be written as the gradient of a scalar function that is proportional to the phase of the wave function condensate.

The hydrodynamic equations have the form of mass and momentum conservation laws and neglecting dissipation they are respectively given by
 \begin{equation} \label{mass-cons}
\partial_t \rho + {\rm div} {\bf j} = 0  \ ,
\ee
and 
\be \partial_t  j_i + \partial_k \Pi_{ik} = 0 \,  ,
\ee
where 
\begin{equation}
\Pi_{ik} = \rho_n  v_{ni} v_{nk} + \rho_s v_{si} v_{sk} + P \delta_{ik} \,,
\end{equation}
is the momentum flux density tensor, and $P$ is the pressure of the system.

Since in a superfluid there can be two different motions, beside Eq.~(\ref{mass-cons}) one has a second hydrodynamic equation  describing the irrotational motion of the superfluid component 
\be
\partial_t {\bf v_s} + \nabla \left(\mu + \frac{{\bf v}_s^2}{2}\right) = 0\,,
\ee
which indicates  that a gradient in the chemical potential acts as a force for the superfluid component.

In order to complete the system of equations, the  energy conservation law is also needed
\begin{equation}
\partial_t E + {\rm div} ({\bf Q}) = 0 \,,
\end{equation}
where $E$ is the energy per unit volume  and the energy flux is given by 
\be
{\bf Q} =  \left( \mu + \frac{{\bf v}_s^2}{2} \right){\bf j} + S T {\bf v}_n
+ \rho_n {\bf v}_n ({\bf v}_n \cdot ({\bf v}_n-{\bf v}_s)) \,,
\ee
where $S$ is the entropy.
In the absence of dissipation, entropy is conserved and one has that 
$\partial_t S + {\rm div} (S{\bf v_n}) = 0 $.

In the presence of dissipative processes  there are additional contributions to the hydrodynamic equations that arise from irreversible processes, thus 
\ba\label{j-dis}
\frac{\partial j_i}{\partial t}  + \partial_j(\Pi_{ij}+ \tau_{ij}) &=&0 \,,\\ \label{v-dis}
\frac{\partial {\bf v_s}}{\partial t} + \nabla \left( \mu + \frac{{\bf v_s}^2}{2}+ h\right) &=& 0 \,,\\ \label{E-dis}
 \frac{\partial E}{\partial t} + \nabla \cdot ({\bf Q} + {\bf Q}^\prime) &=&0 \,, 
\ea
where
\be
{\bf Q}^\prime =  {\bf q} + h ( {\bf j} - \rho  {\bf v}_{n}) +  \tau \cdot {\bf v}_{n} \,,
\ee
and  $\tau_{ij}$, $h$ and ${\bf q}$ are small dissipative terms. In this case  entropy is not conserved and the entropy production rate is given by 
\begin{equation}
R = - h \nabla \cdot(\rho_s ( {\bf v_n}-{\bf v_s})) - \tau_{ik} \partial_k v_{n i} -
\frac{1}{T} {\bf q} \cdot \nabla T \,.
\end{equation} 
From the requirement that the dissipative processes induce an increase of the entropy it follows that 
\ba\label{tau}
 \tau_{ij} &=& - \eta\big(\partial_j   v_{ni} + \partial_i v_{nj} - \frac{2}{3} \delta_{ij} \nabla \cdot
 {\bf v}_n \big) -\delta_{ij} \big(\zeta_1 \nabla \cdot (\rho_s({\bf v}_s -  {\bf v}_n)) + \zeta_2 \nabla \cdot {\bf v}_n\big) \,, \\ \label{h}
 h &=& -\zeta_3\nabla \cdot (\rho_s({\bf v}_s -  {\bf v}_n))- \zeta_4 \nabla \cdot {\bf v_n} \,,\\
{\bf q} &=& - \kappa \,\nabla T \,,
\ea
where $\eta$ is the shear viscosity, $\kappa$ is the  thermal conductivity associated with irreversible heat flow,  and  $\zeta_1,\zeta_2,\zeta_3,\zeta_4$ are the four bulk viscosity coefficients. 
According to the Onsager symmetry principle,  the transport coefficients satisfy the relation
$\zeta_ 1 = \zeta_4$, while the requirement of positive entropy production imposes that 
$\kappa, \eta, \zeta_2, \zeta_3$   are positive and that 
$\zeta_ 1^2 \leq \zeta_2 \zeta_3$. 

Due to the presence of  various bulk viscosity coefficients, the hydrodynamic equations are quite involved. From  
Eq.(\ref{tau}) it is clear that $\zeta_2$ plays the same role as the standard bulk viscosity coefficient.    On the other hand, from Eqs.~(\ref{tau}) and (\ref{h}), one can see that 
$\zeta_1$, $\zeta_3$ and $\zeta_4$ provide a coupling between the hydrodynamic equations of the two components. It is the presence of these couplings that makes the  hydrodynamic equations complicated.

However, notice that the force acting on the superfluid component is proportional to the gradient of $h$ and therefore, even in the presence of dissipation, 
 the motion of the superfluid component will be almost everywhere irrotational, with vorticity concentrated in superfluid vortices, see {\it e.g.}~\cite{Feynman2}.

Let us assume that the superfluid component has a constant density. The  dissipative terms associated with $\zeta_1$ and $\zeta_3$  vanish when $\bf v_s - v_n = {\rm const}$.  Therefore, these viscosity terms oppose to the presence of arbitrary,  space dependent, relative motion between the two components. However, the motion of the normal component is not irrotational (due to the presence of the shear viscosity) whereas the superfluid component is irrotational. This means that it is never possible for these dissipative terms to be zero, unless $\zeta_1$ and $\zeta_3$ are zero or unless a superfluid vortex is created.
Finally, the effect of the dissipative term proportional to $\zeta_4$ is to produce a  force on the superfluid component, due to the variation of the velocity of the normal component.

The friction forces due to bulk viscosities  can also be  understood as  drops in
the main driving forces acting on  the normal and superfluid components. These forces  are given by the gradients
of $P$ and $\mu$, respectively. Therefore we can write that  
\begin{eqnarray}
\label{dissP}
P & =& P_{\rm eq} - \zeta_1 {\rm div}(\rho_s ( {\bf v_n}-{\bf v_s})) - \zeta_2 {\rm div}\,{\bf v_n} \ , \\ \label{dissmu}
\mu & = & \mu_{\rm eq} - \zeta_3 \, {\rm div}(\rho_s ( {\bf v_n}-{\bf v_s})) - \zeta_4 {\rm div}\,{\bf v_n} \,,
\end{eqnarray}
where $ P_{\rm eq}$ and $ \mu_{\rm eq}$ are the pressure and chemical potential in the absence of bulk viscosities. We shall use  this interpretation for the computation of the bulk viscosity coefficients.

In the following Sections we will find that in the conformal limit $\zeta_4=\zeta_1=\zeta_2=0$ whilst $\zeta_3\neq0$, meaning that the only dissipative force acting on the normal component is due to the shear viscosity.  This force will be a source of vorticity for the normal component and since $\zeta_3$ is non-zero, it will generate a  force acting on the superfluid component that will tend  to make $\bf v_s -v_n$ constant.

\subsection{Kinetic theory for the superfluid phonons}

At very low temperatures phonons give the leading thermal contribution to all the thermodynamic
properties of the superfluid. In the hydrodynamic regime, phonons also give the leading contribution
to the transport coefficients entering into the two-fluid equations. Khalatnikov
developed the kinetic theory associated to these degrees of freedom that we briefly review~\cite{IntroSupe}.

Let us first note that at higher  temperatures various quasiparticles or collective modes may contribute  to the thermodynamics or to transport phenomena. For superfluid  $^4$He
the energy spectrum reveals the presence of excitations called rotons, that can  be taken into account in the construction of the kinetic theory~\cite{IntroSupe}. For the cold Fermi gas close to the unitary limit,
finite temperature Quantum Monte-Carlo  simulations~\cite{Bulgac:2005pj} and experimental measurements~\cite{Luo:2007zz},   reveal that only at
very low temperatures the thermal spectrum might be well-described by phonons. At higher temperature the spectrum is much more complicated, and in the computation of transport coefficients
one may need to include other different contributions. However, we shall not consider such a high temperature regime. 

In general one can assume that the dispersion law of phonons is given by 
 \be\label{dispersion}
\epsilon_p= c_s p + B p^3  + {\cal O}(p^5)\,,
\ee
where $c_s$ is the speed of the phonon and for systems with a small coefficient of thermal  expansion it is equal to the speed of first sound.

Under the assumption that the leading contribution  is due to phonons, one can compute various thermodynamic quantities starting from the  phonon distribution function $n$.
The entropy density is given by
\be
 S_{ph} =  \frac{1}{6 \pi^2 T^2 }  \int d p\, p^3 n( n +1) \epsilon_p \frac{\partial \epsilon_p}{\partial p}  \,,
\ee
 the number of phonons per unit volume is given by
\be
{\cal N}_{\rm ph} = \int  \frac{d^3 p}{(2 \pi)^3} \, n  \,,
\ee
and the phonon contribution to the total chemical potential is given by 
\be
\mu = \mu_0 + \frac{1}{2 \pi^2} \int n \frac{\partial \epsilon_p}{\partial \rho} p^2 dp \ ,
\ee
where $\mu_0$ is the chemical potential at zero temperature.

In the study of transport phenomena it is necessary to consider the evolution of the out-of-equilibrium
phonon distribution function $n$, which obeys the kinetic equation

\begin{equation}
\label{Boltzman}
\frac{\partial n}{\partial t} + \frac{\partial n}{\partial {\bf r}} \cdot \frac{\partial H}{\partial {\bf p}}
- \frac{\partial n}{\partial {\bf p}} \cdot \frac{\partial H}{\partial {\bf r}} = C[n] \,,
\end{equation}
where $H= \epsilon_p + {\bf p} \cdot {\bf v}_s$ is the phonon Hamiltonian, and $C[n]$ is  the collision integral.

 At equilibrium  phonons follow the Bose-Einstein distribution
\be n_{\rm eq} (\epsilon_p) = \frac{1}{e^{\epsilon_p/T} -1} \,,
\ee
and the collision term vanishes. For small departures from equilibrium  one can  linearize  the collision term  on the deviations
 $\delta n= n- n_{\rm eq}$ and  the transport coefficients can be obtained by solving the kinetic equation
 obeyed by $\delta n$. In general, this is a complicated task, as one has to deal with an integro-differential equation.  For our purposes it is sufficient to obtain an approximated  expression for  the transport coefficients and therefore we shall use the relaxation time approximation (RTA). In the RTA 
the collision term is written as  
\be
\delta C = - \frac{\delta n}{\tau_{\rm rel}} \ ,
\ee
 where
$\tau_{\rm rel}$ is the relaxation time for the  collisional process   that gives the leading contribution
to the transport phenomena one is studying; for the bulk viscosity coefficients,   collisions that change the phonon number.

Using the RTA one can easily obtain the solution for $\delta n$, and 
 the corresponding dissipative fluxes in the hydrodynamic equations.
 The RTA  provides the correct 
parametric dependence  of the various transport coefficients on the relevant scales, although it does not fix 
with accuracy
the numerical factor in front of these quantities. In the present article  we shall content ourselves with the RTA. The reason is that for a system of cold fermionic atoms the coefficient  of cubic order in  the phonon dispersion law has not been  precisely determined,  therefore there is little motivation for achieving a good precision on the numerical factors of the transport coefficients.

\section{Phonon contribution to the bulk viscosity coefficients}\label{Sec-bulkviscosity}

In this Section we compute the bulk viscosity coefficients for a non-relativistic superfluid.
Khalatnikov described two different methods for the evaluation of these quantities,
one based on studying the evolution of the phonon number density ${\cal N}_{\rm ph}$, see Sec.~\ref{1stmeth}, and the other
one based on studying the evolution of the phonon distribution function $n$, see Sec.~\ref{2onmeth}. 
Here we present in detail both methods and show that the first method corresponds to solving the transport equation in the relaxation time approximation, and thus,  it is equivalent to the second method.

We find that for phonons with a linear dispersion law  all the bulk viscosity coefficients vanish, independent of whether  the system is conformal invariant or not. 
Then, we evaluate the bulk viscosity coefficients for phonons with a cubic dispersion law and
express the result in terms of the parameters  $B$ and $c_s$, see Eq.~(\ref{dispersion}).

\subsection{Evaluation of the bulk viscosity coefficient with the first method}
\label{1stmeth}

When a perturbation applied to a superfluid system  determines a change of the  
number of phonons per unit volume, ${\cal N}_{\rm ph}$,  collisional
processes  tend to restore the equilibrium value of this quantity.
The evolution equation for the phonon number can be written as
\be
\label{Phdenev}
\partial_t {\cal N}_{\rm ph} + {\rm div} ({\cal N}_{\rm ph} {\bf v_n}) = - \frac{\Gamma_{\rm ph}}{T} \mu_{\rm ph} \ ,
\ee
where the rate of change is expressed as a power expansion in a ``fake" phonon chemical
potential, $\mu_{\rm ph}$,  and the decay rate of phonon changing processes, $\Gamma_{\rm ph}$
\cite{footnote-0}.  Expressing the phonon number
 as a function of the density and of entropy, and  
using the linearized continuity equations for these quantities, one can express the phonon chemical potential
in terms of the different dissipative flows that appear in the hydrodynamic equations. These terms modify  the equilibrium pressure and chemical potential, and comparing the results with the expression  in  Eqs.~(\ref{dissP}) and~(\ref{dissmu}), one identifies the different bulk viscosity coefficients.

For small departures from equilibrium and for small values of ${\bf v}_s$ and  ${\bf v}_n$ it turns out 
that~\cite{IntroSupe}
\ba\label{xi1}
\zeta_1 &=& - \frac{T}{\Gamma_{\rm ph}} \frac{\partial {\cal N}_{\rm ph}}{\partial \rho}\left({\cal N}_{\rm ph} -   S \frac{\partial {\cal N}_{\rm ph}}{\partial S} - \rho \frac{\partial {\cal N}_{\rm ph}}{\partial \rho}\right) =  - \frac{T}{\Gamma_{\rm ph}} I_1 I_2 \,,
\\ \label{xi2}
\zeta_2 &=&   \frac{T}{\Gamma_{\rm ph}} \left({\cal N}_{\rm ph} -   S \frac{\partial {\cal N}_{\rm ph}}{\partial S} - \rho \frac{\partial {\cal N}_{\rm ph}}{\partial \rho}\right)^2 =  \frac{T}{\Gamma_{\rm ph}} I_2^2 \,,\\ \label{xi3}
\zeta_3 &=&  \frac{T}{\Gamma_{\rm ph}} \left( \frac{\partial {\cal N}_{\rm ph}}{\partial \rho} \right)^2 =  \frac{T}{\Gamma_{\rm ph}} I_1^2 \,,
\ea
and therefore  $\zeta_1^2 = \zeta_2 \zeta_3$, meaning that one of the relation for positive entropy production is saturated.  Here, we have defined the quantity
\be\label{I1}
I_1=\frac{\partial {\cal N}_{\rm ph}}{\partial \rho} \ ,
\ee
while
\be\label{I2}
I_2={\cal N}_{\rm ph} - S \frac{\partial {\cal N}_{\rm ph}}{\partial S} - \rho \frac{\partial {\cal N}_{\rm ph}}{\partial \rho} \, ,
\ee
and  in order to evaluate the various derivatives that appear in these expressions  we  change variables. Consider that in  Eqs.~(\ref{I1}) and (\ref{I2}) it is assumed that the independent variables are $S$ and $\rho$. Now, we write $S=S(T,\mu_0)$ and $\rho=\rho(T,\mu_0)$, and by the  chain-rule we have that
\ba\label{dn-ds}
\frac{\partial {\cal N}_{\rm ph}}{\partial S} &=& \frac{\partial {\cal N}_{\rm ph}}{\partial T}\frac{\partial T}{\partial S} + \frac{\partial {\cal N}_{ph}}{\partial \mu_0}\frac{\partial \mu_0}{\partial S} \,,\\
\frac{\partial {\cal N}_{ph}}{\partial \rho} &=& \frac{\partial {\cal N}_{ph}}{\partial T}\frac{\partial T}{\partial \rho} + \frac{\partial {\cal N}_{ph}}{\partial \mu_0}\frac{\partial \mu_0}{\partial \rho} \,.\label{dn-drho}
\ea

One can simplify these expressions

 using the Maxwell relation 
\be\label{maxwell}
\left(\frac{\partial T}{\partial \rho}\right)_S 
=\left(\frac{\partial \mu}{\partial S}\right)_\rho  \,,
\ee
and it is now easy to check that with a linear dispersion law all bulk viscosity coefficients vanish.
In order to evaluate  the leading correction in  $B$ to the viscosity coefficients
we define the adimensional parameter
\be 
\label{x-variable}
x = \frac{BT^2}{c_s^3} \,,
\ee
and expand the various quantities evaluated with the equilibrium phonon distribution function to the leading order in $x$.
In this way we obtain that the number of phonons per unit volume is given by
\be
{\cal N}_{ph}  = \frac{T^3 }{2\pi^2 c_s^3}\Big(\Gamma(3) \zeta(3)- x\,  \Gamma(6)\zeta(5) + {\cal O}(x^2)\Big)\,,
\ee
while the  entropy  turns out to be
\be\label{entropy-exp}
 S_{ph} =   \frac{T^3}{6 \pi^2 c_s^3}\Big(\Gamma(5) \zeta(4) -3 x \Gamma(7)\zeta(6)+  {\cal O}(x^2)\Big) \,,
\ee
and the chemical potential is given by
\be
\mu = \mu_0 + \frac{T^4}{2 \pi^2 c_s^4}\left( \frac{\partial c_s}{\partial \rho} \Gamma(4)\zeta(4) + x \Gamma(6)\zeta(6)\Big(\frac{c_s}{B} \frac{\partial B}{\partial \rho}-6 \frac{\partial c_s}{\partial \rho}\Big) +{\cal O}(x^2)\right)  \,.
\ee
In all the above expressions $\Gamma(z)$ and $\zeta(z)$ stand for the Gamma and Riemann zeta functions, respectively.

From these expressions we have that the first non-vanishing corrections to $I_1$ and $I_2$ are 
\be\label{I1-ext}
I_1 = \frac{60}{7 c_s^7 \pi^2} 
 T^5 \Big(\pi^2 \zeta(3) - 7 \zeta(5)\Big) \left(c_s \frac{\partial B}{\partial \rho} - B \frac{\partial c_s}{\partial \rho}\right) \,,
\ee
and
\be\label{I2-ext}
I_2 = -\frac{40  B c_s}{7 c_s^7 \pi^2} 
 T^5 \Big(\pi^2 \zeta(3) - 7 \zeta(5)\Big) - \rho I_1=-\frac{20}{7 c_s^7 \pi^2} 
 T^5 \Big(\pi^2 \zeta(3) - 7 \zeta(5)\Big) \left(2 B c_s + 3 \rho \left(c_s \frac{\partial B}{\partial \rho} - B \frac{\partial c_s}{\partial \rho}\right)\right)\,.
\ee

These expressions are fully general, valid for any non-relativistic superfluid to the leading order in $x$.
Notice that for the next-to-leading order temperature corrections one has to include  terms in the phonon dispersion law going as $p^5$, that we have neglected. In this paper we will only consider
 the first non-vanishing correction to the bulk viscosity coefficients, the 
next-to-leading order corrections being very suppressed at low temperatures.

Once the explicit expressions  of $c_s$ and $B$ and their dependence on the density are known, $I_1$ and $I_2$ can be evaluated.
The quantity that remains to be evaluated is  the decay rate of phonon changing processes, $\Gamma_{\rm ph}$.
The parameter $B$  determines whether  some processes are or are not kinematically allowed.  For $B>0$ the leading 
contribution comes from the Beliaev process  $\phi \to \phi \phi$.
In the opposite  case one has to consider processes like 
$\phi \phi \to \phi \phi \phi$ \cite{footnote}.

\subsection{Evaluation of the bulk viscosity coefficients with the second method}
\label{2onmeth}

The transport coefficients that appear in the superfluid equations 
can be derived by studying the evolution of the deviations from equilibrium of the phonon
distribution function, $\delta n$.  
The Boltzmann equation (\ref{Boltzman}),
amended with a collision term describing phonon number changing processes, is linearized to get the
 equation obeyed by $\delta n$. 
At this stage one could use, for example, the Chapman-Enskog procedure, see e.g.~\cite{Kirkpatrick}. This consists in assuming
that $\delta n$ can be expressed as a function of the hydrodynamical variables and their grandients.
Then, the equation could be solved by a variety of numerical approaches (variational method, use
of orthonormal polynomials, etc). At a technical level, the problem is very much simplified with the
use of the RTA. Then one gets a simple analytical solution for $\delta n$, which is good enough to
get an approximated value of the transport coefficients. The RTA allows us to get the correct
parametric dependence of the transport coefficients on the scales of the problem, but it does
not fix with accuracy the numerical factor in front of these quantities.

Here we show that using the RTA we obtain  the same results found in the previous Section if we take the relaxation time as
\be
\label{rel-time}
\frac{1}{\tau_{\rm rel}} \sim \frac{\Gamma_{\rm ph}}{\cal N_{\rm ph}} \,,
\ee
where $\Gamma_{\rm ph}$ is
the phonon decay rate  appearing in Eq.~(\ref{Phdenev}).

With the use of the RTA one finds the following expressions for the bulk viscosity
coefficients~\cite{IntroSupe} 
\ba\label{second-xi1}
 \zeta_1 &=& - \tau_{\rm rel} \int \frac{d^3 p}{(2 \pi)^3}  \frac{n_{\rm eq}(\epsilon_p) (1+n_{\rm eq}(\epsilon_p))}{T} J_1 \left(\frac 13 {\bf p} \cdot \frac{\partial \epsilon_p}{\partial {\bf p}} + \rho\frac{\partial \epsilon_p}{\partial \rho}\right) \,,\\
 \zeta_2 &=& -\tau_{\rm rel} \int \frac{d^3 p}{(2 \pi)^3}  \frac{n_{\rm eq}(\epsilon_p) (1+n_{\rm eq}(\epsilon_p))}{T} J_2 \left(\frac 13  {\bf p} \cdot \frac{\partial \epsilon_p}{\partial {\bf p}} +   \rho\frac{\partial \epsilon_p}{\partial \rho}\right) \,,\\
\zeta_{3} &=& - \tau_{\rm rel} \int \frac{d^3 p}{(2 \pi)^3}  \frac{n_{\rm eq}(\epsilon_p) (1+n_{\rm eq}(\epsilon_p))}{T} J_1 \frac{\partial \epsilon_p}{\partial \rho} \,,\\
\zeta_4 &=& -\tau_{\rm rel} \int \frac{d^3 p}{(2 \pi)^3}  \frac{n_{\rm eq}(\epsilon_p) (1+n_{\rm eq}(\epsilon_p))}{T} J_2 \frac{\partial \epsilon_p}{\partial \rho} \,,
\ea
where we have defined the quantities
\ba\label{J1}
J_1 &=& \frac{1}{T}\frac{\partial T}{\partial \rho} \epsilon_p -  \frac{\partial \epsilon_p}{\partial \rho}
\,,
 \\ \label{J2}
J_2 &=& \rho J_1 +  \frac{S}{T}\frac{\partial T}{\partial S} \epsilon_p -  \frac{1}3 \frac{\partial \epsilon_p}{\partial {\bf p}}\cdot{\bf p} \,.
\ea

At the leading order in $B$ all the integrals are zero, so we evaluate $J_1$ and $J_2$ at order $B^2$. Then we
realize that they can be written as
\ba
J_1 &=& \left( c_s \frac{\partial B}{\partial \rho} - B \frac{\partial c_s}{\partial \rho}\right) p \frac{M}{L} \ , \\
J_2 &=&  \left(2 B c_s + 3 \rho \left(c_s \frac{\partial B}{\partial \rho} - B \frac{\partial c_s}{\partial \rho}\right)\right)    p \frac{M}{3L}  \,,
\ea
where we have defined the quantities
\ba
M &=& -7 c_s^6 p^2 + 120 \pi^2 B c_s^3 p^2 T^2 + 20 c_s^4 \pi^2 T^2 -896 \pi^4 B c_s T^4 + {\cal O}(B^2) \,,\\
L &=& 7 c_s^6 - 100 B c_s^4 \pi^2 T^2 + {\cal O}(B^2)\,.
\ea

After performing the final integral in momenta, we find that the  expressions of the bulk viscosities obtained by the two methods have the same parametric dependence on physical quantities. While this is what one could
naturally expect, this result only comes out after several subtle cancellations of different terms in the expansion.

Although we are not interested in precise numerical factors, we notice that taking 
\be
\frac{1}{\tau_{\rm rel}} \simeq  7.6 \frac{\Gamma_{\rm ph}}{\cal N_{\rm ph}} \,,
\ee
 the two methods give  the same numerical expressions for the bulk viscosities. In this way, we
check that the first method presented here is  equivalent to solving the transport
equation in the RTA.

\section{Low energy Effective Field theory for the Goldstone mode}\label{Sec-lowenergy}

The information required for the  evaluation of the bulk viscosity coefficients
might be extracted from the effective field theory associated to the Goldstone mode
of the superfluid system. The effective field theory is  constructed as an expansion
over derivatives and over Goldstone fields, and it provides a systematic way to compute
different physical quantities to a given accuracy.

It has been known for a while that the leading order term Lagrangian of the Goldstone
mode of a superfluid system is entirely fixed by  the equation of state~\cite{Popov,Greiter:1989qb}. For the unitarity Fermi gas, the next-to-leading Lagrangian
has been constructed in Ref.~\cite{Son:2005rv}, requiring that it is invariant with respect to non-relativistic general coordinate  and conformal transformations.  We review here the low energy effective field theory
for the cold Fermi system,  
and extract from it the parameters required for the evaluation of the bulk viscosity coefficients,
both in the exact unitarity limit (Sec.~\ref{Ex-EffeThe}), and close to the unitarity limit 
(Sec.~\ref{Close-EffeThe}). We would like to stress that the  methodology used here
might be also followed to study other superfluid systems, governed by different equations of
state, and with different global symmetries.

\subsection{Effective Lagrangian in the exact conformal limit}
\label{Ex-EffeThe}

In the unitarity limit, the thermodynamic properties of the cold Fermi gas can be determined
up to some dimensionless constants \cite{Ho:2004zza}. At  zero temperature, and
due to the absence of any internal scale, dimensional analysis fixes the form of the
pressure as being proportional to that of a  free system
\be
P = c_0 m^4 \mu_0^{5/2} \,,
\ee
where $c_0$ is a dimensionless and universal constant.
This parameter can  be expressed as
\be
c_0 = \frac{2^{5/2}}{15 \pi^2 \xi^{3/2}}\,,
\ee
where  $\xi$ is the universal constant  that fixes the relation between the chemical potential and the Fermi energy $\mu_0 = \xi E_F/m$. Experiments with cold trapped fermionic atoms~\cite{Bartenstein:2004zza} find $\xi \sim 0.32 -0.44$, a result that is in agreement with 
Quantum Monte-Carlo calculations at vanishing temperature~\cite{Chang:2004zz,Astrakharchik:2004zz}. 

The density at zero temperature is then easily deduced 
\be
\rho_0 = \frac 52 c_0 m^4 \mu_0^{3/2} \,,
\ee
and the speed of sound at $T=0$ turns out to be
\be
\label{sp-sound}
c_s = \sqrt{ \rho_0 \frac{\partial \mu_0}{\partial \rho_0}} = \sqrt{ \frac{2 \mu_0}{3}} \ .
\ee

The leading order (LO) Lagrangian for the Goldstone field
can be determined  by the equation of state; the reason being that
 the effective action of the theory at its minimum for constant classical field configurations
has to be equal to the pressure. The next-to-leading order (NLO) Lagrangian is constructed
by demanding invariance with respect to non-relativistic general coordinate invariance and conformal invariance~\cite{Son:2005rv}. The combined terms then read
\be
\label{effL-LO+NLO}
{\cal L} = {\cal L}_{\rm LO}+{\cal L}_{\rm NLO} =c_0 m^{3/2} X^{5/2} +  c_1 m^{1/2} \frac{(\nabla X)^2}{\sqrt{X}} + \frac{c_2}{\sqrt{m}}(\nabla^2 \phi)^2  \sqrt{X} \,,
\ee
where 
\be
X = m \mu_0 - \partial_0 \phi - \frac{({\bf \nabla} \phi)^2}{2 m} \,,
\ee
and $\phi$ is the phase of the condensate and we have neglected effects due to the  trapping potential. Notice that  $\mu_0$ is a chemical potential with dimensions of velocity squared. Our definition of chemical potential is the one given by Khalatnikov in Ref.~\cite{IntroSupe} and differs from the definition given in Ref.~\cite{Son:2005rv} by a mass factor.
Notice that at this order the effective Lagrangian 
depends on two more dimensionless parameters
$c_1$ and $c_2$, that are universal constants.

If one expands the effective Lagrangian in Eq.~(\ref{effL-LO+NLO}) in the field $\phi$ up to quadratic order, and brings the kinetic term into standard form,  one  obtains that the  phonon dispersion relation up to cubic powers of momentum is given by
\cite{Son:2005rv}
\be\label{dispersion-conf}
\epsilon_p =  c_s\left(p  - \pi^2 \sqrt{2 \xi}\left(c_1+ \frac{3}{2} c_2\right) \frac{p^3}{k_F^2}\right)\,,\ee
where $k_F$ is the Fermi momentum, $E_F = k^2_F/2 m$. Then, the speed of the phonon agrees with the
speed of sound, Eq.~(\ref{sp-sound}), and 
we can identify the value of the parameter $B$ introduced in Eq.~(\ref{dispersion}) as 
\be\label{B-conformal}
 B = - \pi^2 c_s \sqrt{2 \xi}\left(c_1+ \frac{3}{2} c_2\right) \frac{1}{k_F^2}\,.\ee

The two parameters $c_1, c_2$ are related to the momentum dependence of the static density and transverse response functions and can be evaluated  by the $\epsilon$-expansion \cite{Rupak:2008xq},
finding $c_1 \simeq -0.0209$, and $c_2/c_1 = {\cal O}(\epsilon^2)$.
These parameters can also be evaluated within mean-field theory \cite{valle},
finding a value for $c_1$ that differs from the previous one by a $30\%$. While the numerical discrepancy between the predictions of the two methods is large, both  give 
$c_1+ \frac{3}{2} c_2 <0$, meaning that $B$ is positive and the Beliaev process $\phi \to \phi \phi$ is kinematically allowed. We will assume that the sign of $B$ is correctly predicted by the two methods, but leave $c_1$ and
$c_2$  as  coefficients still to be determined. 

The three-phonon self-coupling is also determined by expanding the Lagrangian above, and reads  \cite{Rupak:2007vp}
\be
\label{3phLag}
{\cal L}_{3 \phi} =  - \alpha \left(  
 (\partial_0 \phi)^3 - 9 c_s^2 (\partial_0 \phi) ({\bf \nabla} \phi)^2
  \right) + \cdots\,,
\ee
where  terms with higher number of derivatives have been neglected, and the coupling is given by
\be
\label{3-coupling}
 \alpha = \frac{\pi c_s^{3/2} \xi^{3/4}}{3^{1/4} 8 \, m^2 \mu_0^2} \ .
 \ee
 
The effective field theory here presented allows us  to compute the  corrections to the
phonon dispersion law needed in our computations. We have checked that the one-loop corrections
to the dispersion law  are  very suppressed. Further, they do not  modify the
sign of the coefficient $B$ that appears at tree level, which is relevant in deciding
whether the Beliaev decay is kinematically allowed or not.
 A more detailed discussion of the one-loop phonon dispersion law will be presented elsewhere.

\subsection{Effective Lagrangian close to the conformal limit}
\label{Close-EffeThe}

For  finite scattering length, the pressure can be written as a power expansion in $1/a$
\be
P = P_0 + P_{\rm CB} = c_0  m^4 \mu_0^{5/2} + \frac{d_0 m^3 \mu_0^2}{a} + \cdots \,,
\ee
where  $d_0$ is a dimensionless constant to be determined by matching or experimentally.

This change in the equation of state  induces new terms in the effective Lagrangian
which are responsible for the breaking of scale and  conformal invariance \cite{Son:2005rv}
\be\label{L-cb}
{\cal L} = {\cal L}_{\rm LO} +{\cal L}_{\rm NLO}+ {\cal L}_{\rm CB} + \cdots = c_0 m^{3/2} X^{5/2} +  c_1 m^{1/2} \frac{(\nabla X)^2}{\sqrt{X}} + \frac{c_2}{\sqrt{m}}(\nabla^2 \phi)^2  \sqrt{X} + d_0 \frac{m X^2}{a} + \cdots
\ee

The effect of the scale breaking term is to change the  phonon 
dispersion law, and the self-coupling coefficients. In order to find the proper transformation of these quantities we
expand ${\cal L}$ to quadratic order in $\phi$, and
 bring  the kinetic term  into the canonical form by  the field rescaling
\be
\phi  \rightarrow  \phi \, \frac{\pi \xi^{3/4}}{(2 \mu_0)^{1/4} m} \left(1 - y\right) \,,    
\ee
where we have defined the quantity
\be
\label{par-breaking}
y \equiv \frac {d_0 \pi^2\xi^{3/2}} { a m\sqrt{2 \mu_0} } \,.
\ee
One then finds that the speed of the phonon is corrected at the leading order in $1/a$ as
\be
\label{cf-speed}
c_{s,{\rm CB}}  = c_s \left(1 + \frac y2  \right) \,,
\ee
while the cubic term in the phonon dispersion law is now given by
\be
\label{cf-B}
 B_{\rm CB} =  - \frac{\pi^2 \xi^{3/2}}{\sqrt{3 \mu_0} m^2 }  \left( 
c_1 \left(1 - \frac{3 y}{2  } \right)+\frac{3}{2}c_2  \left(1 - \frac{5  y}{ 2 } \right) \right)\,.
\ee

The three-phonon self-coupling is also modified as
\be
{\cal L}_{\rm CB}^{3 \phi} = - \alpha_{\rm CB}\left[(\partial_0\phi)^3-9  c_{s,{\rm CB}}^2\left(1+ y \right)(\partial_0\phi)({\bf\nabla}\phi)^2\right] + \cdots
\ee
where the coupling reads
\be
\alpha_{\rm CB} =\alpha\left(1- 3 y \right) \,.
\ee 

The density at zero temperature of the system also deviates from its value in the exact unitarity limit
by

\be
 \rho^{\rm CB}_0 =\rho_0 \left(1+ 3 y\right) \,.
\ee

These corrections to the  parameters of the phonon dispersion law and to the  self-couplings coefficients will turn out 
to be of great importance in the  computations of the bulk viscosity coefficients, because they will allow us to compute the only non-vanishing contributions to $\zeta_1$ and $\zeta_2$ close to the conformal limit.

\section{Bulk viscosities  close to the unitarity limit}
\label{Sec-bulkviscosity-CA}

For the explicit evaluation of the bulk viscosity coefficients, we  start by considering the exact unitarity limit. Using the expressions of $c_s$ and $B$ 
given respectively in Eqs.~(\ref{sp-sound}) and (\ref{B-conformal}), we evaluate the derivatives with respect to the
density. At low temperature we can approximate

 \be\label{B-derivative}
\rho \frac{\partial B}{\partial \rho} \approx \rho_0 \frac{\partial B}{\partial \rho_0} = \rho_0 \frac{\partial B}{\partial \mu_0}   \frac{\partial \mu_0}{\partial \rho_0}  =  -\frac{B}{3}\,,\ee 
where  we have used the fact that $\frac{\partial \mu_0}{\partial \rho_0} = \frac{2 \mu_0}{3\rho_0}$.  In a similar way we find  that at unitarity and for  vanishing temperatures
\be\label{cs-derivative}
 \rho \frac{\partial c_s}{\partial \rho} \approx \frac{c_s}{3}\,.
\ee

Upon substituting the expressions in Eqs.~(\ref{B-derivative}) and~(\ref{cs-derivative}) in Eqs.~(\ref{I1-ext}) and (\ref{I2-ext})  it  follows that
\be\label{I2conf}
 I_1 =-\frac{40  B }{7 c_s^6 \pi^2 \rho_0} 
 T^5 \Big(\pi^2 \zeta(3) - 7 \zeta(5)\Big) \, \ee
and
  \be
I_2 =0  \,.
\ee
Therefore the bulk viscosity coefficients $\zeta_1$ and $\zeta_2$ vanish. 

To obtain the final expression for $\zeta_3$ we still have to evaluate the decay rate $\Gamma_{\rm ph}$.  If  the process $\phi \rightarrow \phi \phi$  is kinematically allowed, as suggested by the results of
Refs.~\cite{Rupak:2008xq,valle}, 
 then we have that
\be
\Gamma_{\rm ph} =  \int \frac{d^3 p}{(2 \pi)^3} \frac{d^3 q}{ (2 \pi)^3} 
\frac{d^3 k}{ (2 \pi)^3} | { M}|^2 n_{\rm eq}(\epsilon_p) \left(1 +  n_{\rm eq}(\epsilon_q)\right)\left(1 +  n_{\rm eq}(\epsilon_k)\right) (2 \pi)^4 \delta^{(4)} (P-K-Q) \ ,
\ee 
where the delta functions ensure the energy-momentum conservation, where  $P  = (\epsilon_p, {\bf p})$, $Q  = (\epsilon_q, {\bf q})$, $K  = (\epsilon_k, {\bf k})$ and  
where $|M(P,Q,K)|^2$ is the squared of the scattering amplitude with non-relativistic normalization of the integration measure. This quantity is
 related to the squared of the scattering matrix with relativistic conventions $| {\cal M}|^2$ by 
\be
|M(P,Q,K)|^2 = \frac{| {\cal M}(P,Q,K)|^2}{ 2 \epsilon_p 2 \epsilon_q 2 \epsilon_k} \,,
\ee
and can be computed from the three-phonon interaction, 
 Eq.~(\ref{3phLag}).
The scattering matrix takes a particular simple form if expressed in terms
of the vector product defined as $P \cdot K = \epsilon_p \epsilon_k - 9 c_s^2 {\bf p}\cdot {\bf k}$ 
\cite{Rupak:2007vp}
\be
| {\cal M}(P,Q,K)|^2 = 4 \alpha^2 \left(\epsilon_p \, Q \cdot K + \epsilon_q \, P \cdot K +\epsilon_k 
\, Q \cdot P \right)^2 \,,
\ee
where
$\alpha$ is defined in Eq.~(\ref{3-coupling}).

For phonons with a linear dispersion law, the splitting process is perfectly collinear. For $B>0$  the process is non-collinear, and the angle of
scattering between the incoming phonon and the two outcoming phonons is proportional to $B$.
Since $I_1$ and $I_2$ are proportional to $B$ we can evaluate the decay width considering the collinear process only and  from the expressions above  we obtain that 
\be
\Gamma_{\rm ph} \simeq 14781.6 \frac{\alpha^2 T^8}{\pi^3  c_s^6}  + {\cal O} (x) \,,
\ee 
where $x$ is defined in Eq.~(\ref{x-variable}). Note that in the exact unitarity limit $x \sim (T/\mu_0)^2$, so that
at low temperature it is a good approximation to neglect terms of order $x$ in $\Gamma_{\rm ph}$.

Upon substituting  the decay width and the expression for $I_1$ reported in Eq.~(\ref{I2conf}) in Eq.~(\ref{xi3}) we find that
\be
\zeta_3  \simeq  0.015 \frac{B^2 T^3}{c_s^6 \alpha^2 \rho_0^2} + {\cal O}\left(T^5\right) \simeq  3695.4 \left( \frac \xi \mu_0 \right)^{9/2} \frac{(c_1 + \frac 32 c_2)^2}{m^8} T^3 + {\cal O}\left(T^5\right) \,.
\ee

It is worth noticing that if the coefficient $B$ were negative, then one should consider the decay
rate associated to the process $\phi \phi \to \phi \phi \phi$, meaning that one would get a different
temperature dependence for 
 $\Gamma_{\rm ph}$, and thus, also for $\zeta_3$.

With the low energy effective field theory, one can as well compute the first corrections to the
bulk viscosity coefficients for finite  $s$-wave scattering length $a$.
We have to evaluate how all the various ingredients needed in the computation are corrected
in the presence of this scale breaking effect. 

The decay rate is modified as
\be
\Gamma_{\rm ph}^{\rm CB} \simeq  \Gamma_{\rm ph} \left( 1 - \frac{27 }{4 } y\right ) + {\cal O}(y^2) \,,
\ee
where $y$ has been defined in Eq.~(\ref{par-breaking}).

The expressions of $I_1$ and $I_2$ are also affected and we find that
 \ba
I_1^{\rm CB} & = & I_1 \left(1 - \frac{15 c_1 + 27 c_2}{2 c_1 + 3 c_2} y\right) + {\cal O}(y^2)\,, \\
\label{cf-I2}
I_{2}^{\rm CB} & = &
\frac{20}{7 c_s^7 \pi^2} 
 T^5 \Big(\pi^2 \zeta(3) - 7 \zeta(5)\Big) \frac{\pi^2 \xi^{3/2} c_2}{\sqrt{2} m^2} y + {\cal O}(y^2)\,. 
\ea

Upon substituting these expressions in Eqs.(\ref{xi1}), (\ref{xi2}) and (\ref{xi3}) we find that to
leading order in $y \sim 1/a$
\ba
\zeta_1 &\simeq& -264.7 \, c_2 \left(c_1 +\frac{3}2 c_2\right)\frac{T^3\xi^3}{m^4 \mu_0^3} y \,, \\
\zeta_2 &\simeq &19.0 \, c_2^2 \frac{T^3 \xi^{3/2}}{\mu_0^{3/2}}y^2 \,, \\
\zeta_3   &\simeq & 3695.4 \left( \frac \xi \mu_0 \right)^{9/2} \frac{(c_1 + \frac 32 c_2)^2}{m^8} T^3 \left(1 -  \frac{66 c_1 + 135 c_2}{8 c_1 + 12 c_2 } y\right)\,.
\ea
Notice that the first non vanishing correction to $\zeta_2$ is of the order of $1/a^2$, meaning that this bulk viscosity coefficient has to be neglected in the  hydrodynamic equations~(\ref{j-dis}),(\ref{v-dis}) and (\ref{E-dis}), where we are retaining  terms of order $1/a$ only.

\section{Conclusion}\label{Sec-conclusion}

We have  derived the expressions for   the three independent bulk viscosity coefficients  of a non-relativistic Fermi superfluid   close to unitarity.  In doing this we have first derived general expressions for these three  transport coefficients assuming that
the leading  contribution comes from the superfluid phonons, with dispersion law containing both a linear and a cubic term in momentum. Our computations are valid at very low temperatures, because for temperatures close to the superfluid transition the contribution of other quasiparticles might also be important. 

In agreement with the outcome of Ref.~\cite{Son:2005tj} we find by  explicit calculation that at unitarity the bulk viscosity coefficients $\zeta_1$ and $\zeta_2$ vanish, whilst    
\be
\zeta_3 \simeq  3695.4 \, \hbar^4 \left( \frac \xi \mu_0 \right)^{9/2} \frac{(c_1 + \frac 32 c_2)^2}{m^8} (k_B T)^3 \,,
\ee
and we have restored here both the Planck and Boltzmann constants.
In evaluating  $\zeta_3$ we have considered
that the leading collisional process is the phonon decay $\phi \rightarrow \phi \phi$, which  
is kinematically allowed  according to the results 
of Refs.~\cite{Rupak:2008xq,valle}. If this were not the case, this transport coefficient would
be dominated by the process $\phi \phi \rightarrow \phi \phi \phi$, resulting in a very different temperature dependence.

Let us point out that in the computation
of the shear viscosity  for the same system \cite{Rupak:2007vp}, it was considered that the decay
$\phi \rightarrow \phi \phi$ was
not kinematically allowed, resulting in a dependence  $\eta \propto 1/T^5$.
 Considering this decay process for the computation of the shear viscosity, would change
its temperature dependence into  $\eta \propto 1/T$, as it occurs for $^4$He~\cite{Maris}.
We thus realize that it is very important to determine with good precision the phonon dispersion law, as the explicit
values and temperature dependence
of all the transport coefficients are extremely sensitive to its  form.

We have also evaluated corrections to the bulk viscosities due to conformal breaking terms, finding
\ba
\zeta_1 &\simeq& -264.7 \, \hbar \, c_2 \left(c_1 +\frac{3}2 c_2\right)\frac{\xi^3 (k_B T)^3}{m^4 \mu_0^3} y \,,\\
\zeta_2 &\simeq & \frac{19.0}{\hbar^2} \, c_2^2 \frac{ \xi^{3/2}(k_B T)^3}{\mu_0^{3/2}}y^2 \,,\\
\zeta_3   &\simeq & 3695.4 \,\hbar^4 \left( \frac \xi \mu_0 \right)^{9/2} \frac{(c_1 + \frac 32 c_2)^2}{m^8} (k_B T)^3 \left(1 -  \frac{66 c_1 + 135 c_2}{8 c_1 + 12 c_2 } y\right)\,.
\ea
We have restricted the computation to the leading correction in the parameter $y= \hbar  \frac{d_0 \pi^2\xi^{3/2}} { a m\sqrt{2 \mu_0} }$ , which measures the departure from the  conformal limit caused by a large but finite value of the scattering length $a$.

The presence of non-vanishing bulk viscosity coefficients might be experimentally detectable. As an example  it influences the propagation of first and second sound in a superfluid~\cite{IntroSupe}. The damping of first sound, $\alpha_1$,  depends on the shear viscosity  and on $\zeta_2$, whereas the damping of second  sound, $\alpha_2$, depends on all  the dissipative coefficients. Since shear viscosity  is much larger than the others it will give the leading contribution to the sound absorption coefficients.  However,  one can show that in the unitary limit
\be
\frac{\rho_n}{3^{3/2} \rho_s}\alpha_2 - \alpha_1 = \frac{\omega^2}{2 \rho_n c_s^3} \left(\rho^2 \zeta_3 + \frac{\rho_n \kappa}{\rho_s T}\frac{\partial T}{\partial S}\right)
\ee
and therefore this combination of the absorption coefficients is independent of $\eta$. Here $\omega$ is the frequency of the sound oscillation and $\kappa$ is the thermal conductivity. In order to evaluate the thermal conductivity one has to consider processes that change the total momentum of phonons, meaning that one has to consider the scattering of phonons with different (quasi)particles, see {\it e.g.}~\cite{kittel}. A more detailed description of the thermal properties of unitary superfluids will be presented elsewhere.  

It might be interesting to explore the possibility of measuring the bulk viscosities in trapped Fermi superfluids through the study of dipole and breathing modes.
The  two-fluid equations of Landau predict the presence of different hydrodynamic modes, according to whether the
superfluid and normal components oscillate in-phase, or move out-of-phase~\cite{out-of-phase-theory}. It has been suggested
that the out-of-phase oscillations might be experimentally detected~\cite{out-of-phase-exp}. If this is the case, its study might lead to the determination of the transport coefficients studied here, in the same way that the shear
viscosity can be extracted studying the in-phase breathing mode~\cite{Schafer:2007pr}.

\begin{acknowledgments}
This work was supported by the Spanish grant
FPA2007-60275. M.A.E. was also supported by MICCIN FPU (Spain).
\end{acknowledgments}


\begin{thebibliography}{99}
\bibitem{footnotezetas}
The shear and bulk viscosities are also traditionally known as first and second viscosities, respectively,
but we will not use that terminology here.
\bibitem{footnote-0}
Our definition of $\Gamma_{\rm ph}$ differs by a factor $T$ from that used in Ref.~\cite{IntroSupe}.
\end{thebibliography}
\end{document}